\documentclass[10pt,a4paper,twocolumn]{article}
\usepackage{hyperref}
\usepackage[margin=0.9in]{geometry}
\usepackage[version=3]{mhchem}
\usepackage{authblk}
\usepackage[version=3]{mhchem}
\usepackage[square,numbers,sort&compress]{natbib}
\usepackage{amsmath}
\usepackage{amssymb}
\usepackage{amsthm}
\usepackage{graphicx}
\usepackage{balance}
\usepackage{abstract}

\title{Formation, dissolution and properties of surface nanobubbles}
\author[1]{Zhizhao Che\thanks{chezhizhao@tju.edu.cn}}
\author[2]{Panagiotis E.\ Theodorakis}
\affil[1]{\small{State Key Laboratory of Engines, Tianjin University, Tianjin, 300072, China.}}
\affil[2]{\small{Institute of Physics, Polish Academy of Sciences, Al.\ Lotnik\'ow 32/46, 02-668 Warsaw, Poland.}}

\date{}

\begin{document}

\twocolumn[
  \begin{@twocolumnfalse}
    \maketitle
    \begin{abstract}
Surface nanobubbles are stable gaseous phases in liquids that form on solid substrates. While their existence has been confirmed, there are many open questions related to their formation and dissolution processes along with their structures and properties, which are difficult to investigate experimentally. To address these issues, we carried out molecular dynamics simulations based on atomistic force fields for systems comprised of water, air (\ce{N2} and \ce{O2}), and a Highly Oriented Pyrolytic Graphite (HOPG) substrate. Our results provide insights into the formation/dissolution mechanisms of nanobubbles and estimates for their density, contact angle, and surface tension. We found that the formation of nanobubbles is driven by an initial nucleation process of air molecules and the subsequent coalescence of the formed air clusters. The clusters form favorably on the substrate, which provides an enhanced stability to the clusters. In contrast, nanobubbles formed in the bulk either move randomly to the substrate and spread or move to the water--air surface and pop immediately. Moreover, nanobubbles consist of a condensed gaseous phase with a surface tension smaller than that of an equivalent system under atmospheric conditions, and contact angles larger than those in the equivalent nanodroplet case. We anticipate that this study will provide useful insights into the physics of nanobubbles and will stimulate further research in the field by using all-atom simulations.
    \end{abstract}
  \end{@twocolumnfalse}
]
\saythanks
%
%
%
\section{Introduction}
\begin{figure}
    \begin{center}
    \includegraphics[scale=0.8]{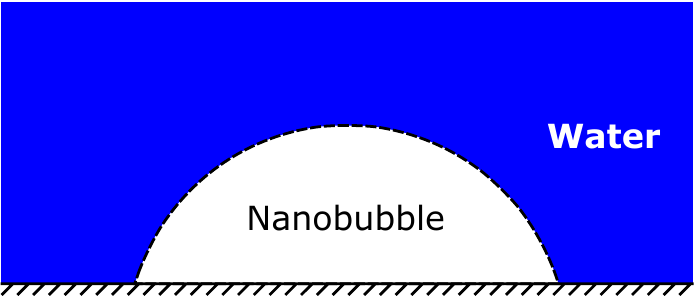}
    \end{center}
    \caption{\label{fig:1} Schematic illustration of a nanobubble (in white) surrounded by water (in blue) on a solid substrate (bottom).}
\end{figure}

Surface nanobubbles are gaseous phases that can form spontaneously at solid--liquid interfaces \cite{Lohse2015,Seddon2011} (Figure~\ref{fig:1}). They were observed in experiments to typically have diameters of 50--100 nm and heights of 10--20 nm.
Moreover, they are unexpectedly stable, namely they can exist for days without dissolving as, for example, in the case of a Highly Oriented Pyrolytic Graphite (HOPG) substrate immersed in water \cite{Lohse2015}. This is only one of the features that have motivated research on nanobubbles in various fields, including the relatively new field of plasmonic bubbles in the context of energy conversion \cite{Lukianova2010,Adleman2009}.
In addition, nanobubbles find important applications in nanomaterial engineering \cite{Huang2009},
transport in nanofluidics (e.g., autonomous motion of nanoparticles) \cite{Bocquet2010},
catalysis and electrolysis \cite{Luo2014}, cleaning \cite{Wu200810cleaning}, and flotation \cite{Calgaroto2014}.
In the case of flotation, for example, nanobubbles attached to nanoparticles of certain sizes rise together due to buoyancy, in this way facilitating the separation of nanoparticles or even fine oil nanodroplets.

The existence of nanobubbles was speculated about twenty years ago \cite{Parker1994},
while the first atomic-force microscopy (AFM) image of nanobubbles was taken in the year 2000 \cite{Lou2000,Ishida2000}.
Initially \cite{Ball2003}, researchers believed that the presence of nanobubbles on the image was an artifact, but the use of additional methods confirmed their existence \cite{Seo2007,Martinez2007,Zhang2008}.
However, the experimental characterization of these nano-objects still remains a difficult task
despite significant efforts over the last decade.
For example, the most popular method for studying nanobubbles, AFM \cite{Lou2000,Ishida2000,Hampton2008AFM}, can only provide long time averages, preventing the study of nanobubble formation and dissolution \cite{Lohse2015}. In another example, surface plasmon resonance spectroscopy is sensitive to the chemical properties of interfaces \cite{Zhang2007}.
As a result, it is difficult to isolate the properties of the nanobubbles from
those of the interface. In addition, the control of the nanobubble size in experiments has to be involved,
which is crucial for analyzing their size-dependent properties \cite{Lou2000,Yang2003}.
These are only a few of the reasons that many universal properties of nanobubbles still remain unexplored \cite{Lohse2015}.
For example, nanobubbles are known to be stable for days,
which contrasts with the expectation of their rapid dissolution within the diffusive time scale (microseconds)
due to the high Laplace pressure inside nanobubbles \cite{Borkent2009}.
Some studies suggest that contamination on the substrate may play a role in this stability \cite{Ducker2009},
whereas other studies explain this phenomenon through the balance between gas influx and outflux \cite{Brenner2008,Weijs2013}, pinning \cite{Liu2014b,Guo2015,Guo2016}, and gas oversaturation \cite{Liu2014,Maheshwari2016}.
Recently, molecular dynamics (MD) simulations of a coarse-grained model suggested that nanobubbles dissolve within
less than a microsecond, claiming that the experimental nanobubbles are stabilized by a nonequilibrium mechanism \cite{Weijs2012}.

Hence, many fundamental questions are still open; even the proof that nanobubbles actually consist of a gaseous phase is under debate \cite{Lohse2015}. Clearly, we need systematic approaches to study the formation, dissolution and properties of nanobubbles targeting realistic systems. Understanding these phenomena requires the access to an atomistic/molecular-scale description of the phenomena, the absolute control of system parameters (e.g., thermodynamic conditions), and the ability to follow in detail the time evolution of the formation and dissolution processes of nanobubbles. To address these issues, we employed MD simulations based on atomistic force fields. In particular, we studied the spontaneous formation and dissolution mechanisms of air nanobubbles (\ce{N2} and \ce{O2} molecules) on an HOPG substrate immersed in water.
This paper describes simulation results for this realistic situation and provides insights into morphological properties of nanobubbles, such as their size, density, contact angle, and interfacial tension.

\section{Materials and methods}\label{sec:MaterialMethod}

\begin{figure}[tb]
    \begin{center}
    \includegraphics[width=\columnwidth]{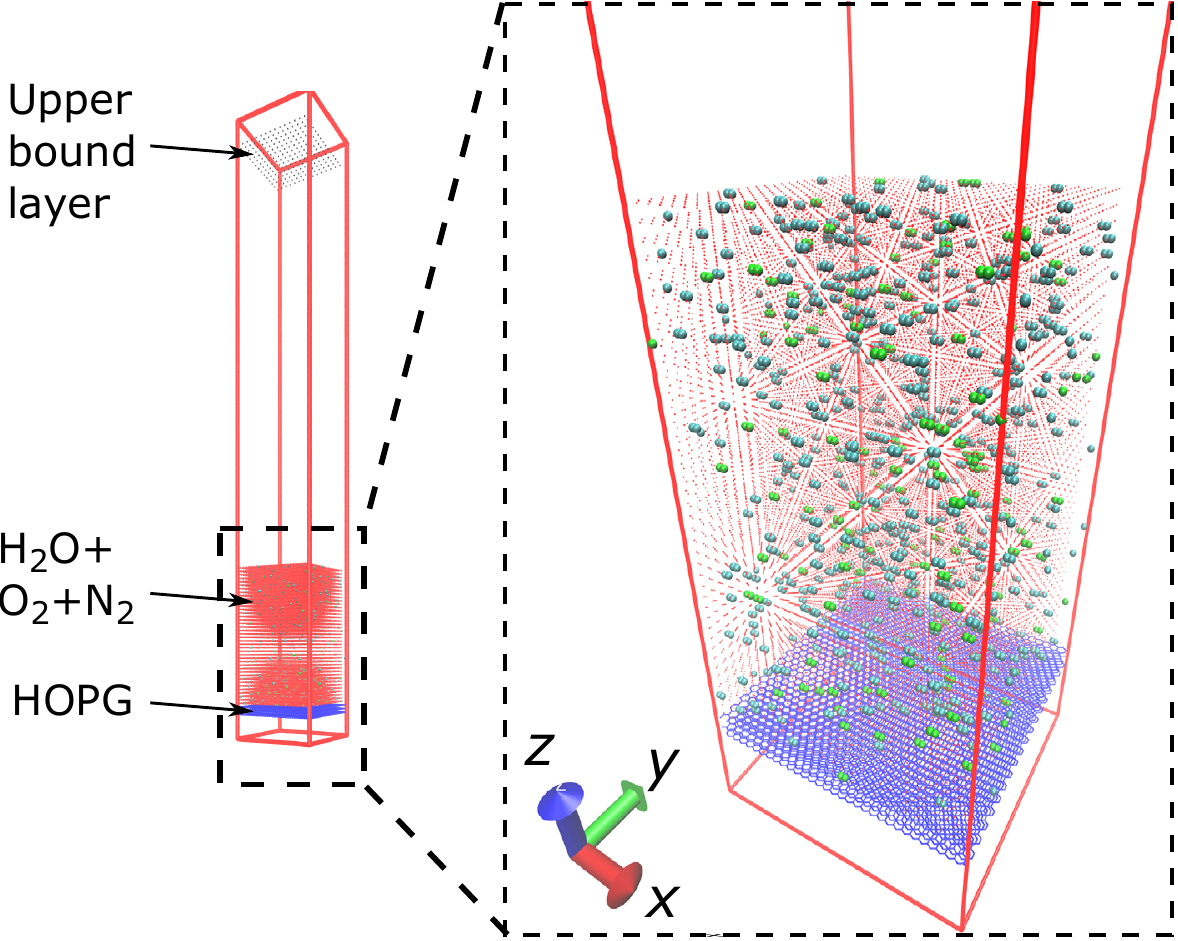}
    \end{center}
    \caption{\label{fig:2} Perspective view of an initial configuration of the system used in a simulation to investigate the subsequent formation of surface nanobubbles on an HOPG substrate in water. The magnified region highlights the gas molecules (cyan color for \ce{N2} and green color for \ce{O2}) dispersed randomly within the aqueous medium (in red). The HOPG substrate (in blue) at the bottom of the simulation box consists of three layers of carbon atoms arranged on a hexagonal lattice. The red straight lines indicate the boundaries of the simulation box with periodic boundary conditions applied in the $x$ and $y$ directions. The top of the simulation box is bound by a silicon wall (in black).}
\end{figure}

\subsection{System configuration}
The system consisted of an HOPG substrate at the bottom of a rectangular simulation box, air (\ce{N2} and \ce{O2}) and water (\ce{H2O}) molecules, and a solid layer of silicon atoms on a square lattice at the top of the simulation box to regulate the pressure in the system and to prevent the air molecules from escaping from the simulation box (see Figure \ref{fig:2}). The top silicon layer and the bottom substrate were normal to the $z$ direction and defined the size of the system in this direction, while periodic boundary conditions were applied in the $x$ and $y$ directions. The distance between the top surface and the water molecules was large enough to guarantee that these two components did not interact directly during the simulations. Moreover, the distance between the top surface and the liquid--vapor interface was large enough to capture the air molecules escaping from the liquid phase.

The amount of water used in experiments is beyond the reach of molecular-level simulations, which makes the simulation of the whole system almost impossible due to the high demand in CPU time. A test of different water volumes was performed and a largest possible system was selected to consider most features of the relevant processes with reasonable requirement in CPU time. Therefore, the escape rate of the air molecules was inevitably much higher than in experiments. However, it is believed that this does not affect significantly the understanding of the formation and the properties of nanobubbles \cite{Weijs2012}.
The HOPG substrate is one of the most well-studied cases in the context of nanobubbles in experiments \cite{Zhang2007a,Zhang2008b,Zhang2009c}, hence our choice in this study. Here, the HOPG substrate is an ideal form of synthetic graphite (without contamination or defects) consisting of several layers of carbon atoms in a hexagonal arrangement (see Figure \ref{fig:2}).

\subsection{MD simulation method}\label{sec:SimDetail}
The study was performed by using MD simulations of atomistic force fields \cite{Berendsen1984,Potoff2001,Zhang2006}.
Simulations were carried out in the NVT ensemble by using the Nos\'{e}-Hoover thermostat as implemented in the LAMMPS package \cite{Plimpton1995}.
Therefore, the number of particles and the volume of the system were constant during the simulations,
while the temperature was allowed to fluctuate around a predefined value, which in our case was 300 K. Henceforth, quantities related to the system were expressed in real units.

The total energy of a system includes harmonic interactions described by the following relations,
\begin{equation}\label{eq:force}
    U=\sum\limits_{\text{bonds}}{{{k}_{b}}}{{(l-{{l}_{0}})}^{2}}+\sum\limits_{\text{angles}}{{{k}_{a}}}{{(\theta -{{\theta }_{0}})}^{2}},
\end{equation}
where ${{k}_{b}}$ and ${{k}_{a}}$ are stiffness parameters for bond and angle interactions, respectively; ${{l}_{0}}$ and ${{\theta }_{0}}$ are the equilibrium distance and the equilibrium angle, while $l$ and $\theta $ are the actual distance and angle during the simulation at a particular time-step. For nonbonded interactions, a pairwise potential consisting of Lennard--Jones (LJ) 12--6 and Coulomb interactions was considered,
\begin{equation}\label{eq:ljc}
    {{U}_{ij}}=4{{\varepsilon }_{ij}}\left[ {{\left( \frac{{{\sigma }_{ij}}}{{{r}_{ij}}} \right)}^{12}}-{{\left( \frac{{{\sigma }_{ij}}}{{{r}_{ij}}} \right)}^{6}} \right]+\frac{{{q}_{i}}{{q}_{j}}}{{{r}_{ij}}},
\end{equation}
where ${{\varepsilon }_{ij}}$ expresses the strength of the interaction, ${{r}_{ij}}$ the actual distance between the $i$th and the $j$th particles, ${{\sigma }_{ij}}$ the size of the particles, and ${{q}_{i}}$ the charge of the $i$th particle. Charges were taken into account by using the particle--particle particle--mesh (PPPM) method \cite{Hockney1989}. The SPC/E model was employed for the simulation of water molecules. The details of the simulations are provided in Supplementary Materials.

\subsection{Properties of nanobubbles}
The nanobubble was determined firstly by cluster analysis from the trajectories of the molecules. Then, Voronoi analysis was performed to find the interface between the water phase and the air phase. The volume, the density, and the contact angle of nanobubbles were calculated based on this interface (See Supplementary Materials for details). The contact angle of water nanodroplets on the HOPG substrate was also analyzed for comparison. The various analysis was performed by using customized Matlab programs.

To calculate the surface tension of nanobubble in water, a new simulation configuration was used by setting up a layer of water molecules in air at the nanobubble density (and a configuration with air at normal density for comparison). For a planar interface between two phases (1 and 2) perpendicular to the $z$ direction, the surface tension $\gamma_{12}$ can be calculated as follows \cite{Shi2006SurfaceTensioni},
\begin{equation}
\gamma_{12} = \int_{\rm phase 1}^{\rm phase 2}[(P_{xx}+P_{yy})/2 - P_{zz}]dz,
\end{equation}
where $P_{xx}$ and $P_{yy}$ are tangential pressure components, $P_{zz}$ the normal pressure component, and the integration is along the normal direction to the interface.

\section{Results and discussion}

\subsection{Formation and dissolution of nanobubbles}
\begin{figure}
    \begin{center}
    \includegraphics[width=\columnwidth]{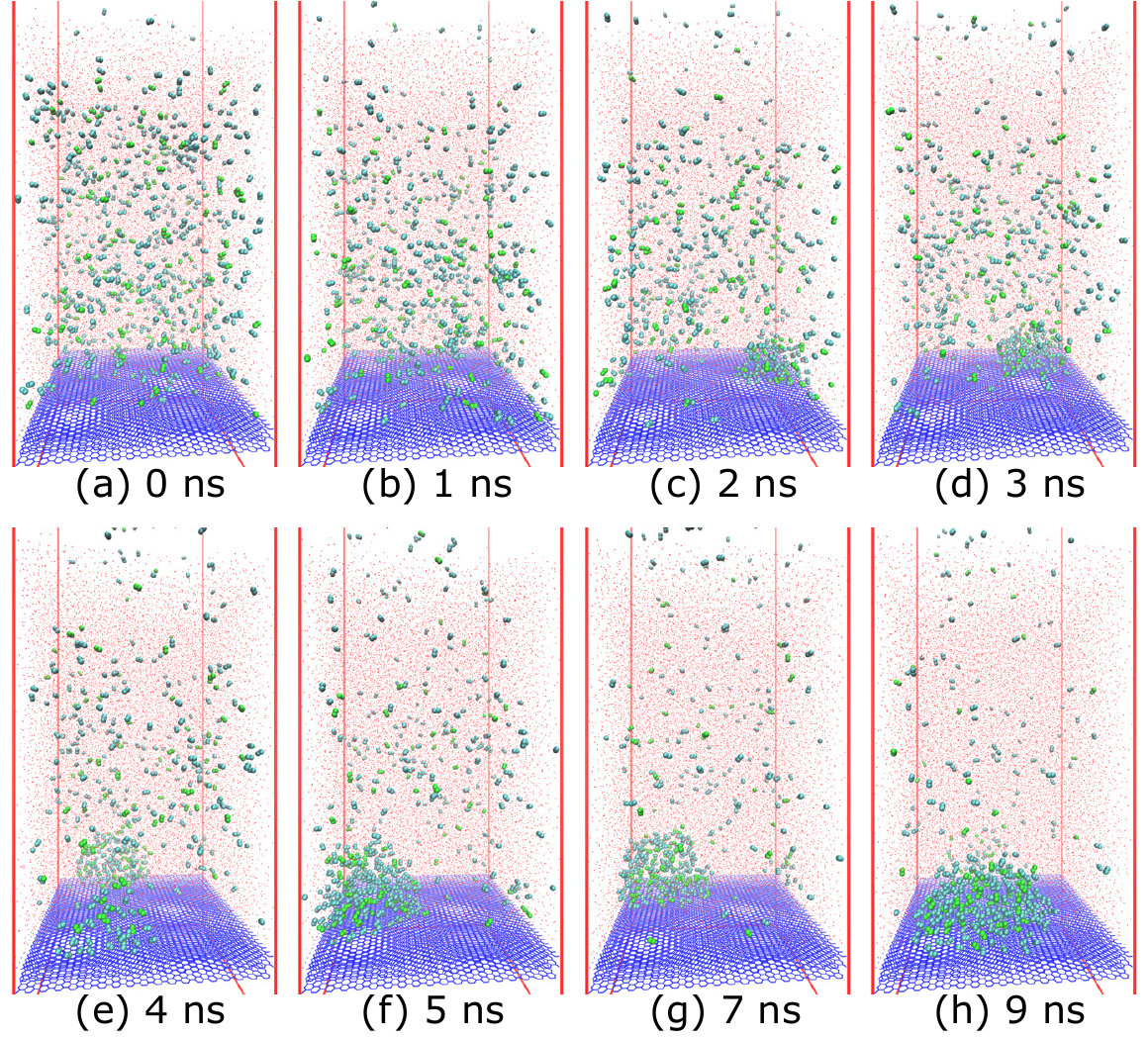}
    \end{center}
    \caption{\label{fig:6formation} Formation of a nanobubble on the HOPG substrate illustrated by snapshots at different times as indicated. The water/air mixture is initially homogeneous, and the air molecules gradually aggregate into small clusters dispersed randomly in water. Then, some clusters collide to form larger ones. Finally, a large cluster form on the substrate due to the interaction with the substrate. The nanobubble exhibits a Brownian-like motion in contact with the substrate. The domain is periodic in the $x$ and $y$ directions. A movie for the whole process is available as Supplementary Material Video 1. }
\end{figure}

The formation of nanobubbles is driven by a nucleation process (see Figure \ref{fig:6formation}).
Initially, air molecules are dispersed randomly and uniformly in water. Later, they aggregate into clusters, which gradually grow with the addition of nearby air molecules. This is due to the oversaturation of the air molecules in water and the favorable interaction among air molecules in comparison with the less favorable interaction between air and water molecules. Moreover, the interaction between air molecules and the HOPG substrate is also more favorable in comparison with the interaction between air and water molecules. As a result, the energetically most suitable position of air molecules within the aqueous phase is on the substrate. Therefore, the air molecules dispersed in water diffuse towards the substrate, eventually forming one or more nanobubbles. Other air molecules, however, may have diffused towards the liquid--vapor interface without interacting with the substrate, in this way escaping from the aqueous phase and eventually reaching the gas phase above the aqueous phase.

\begin{figure}
    \begin{center}
    \includegraphics[width=\columnwidth]{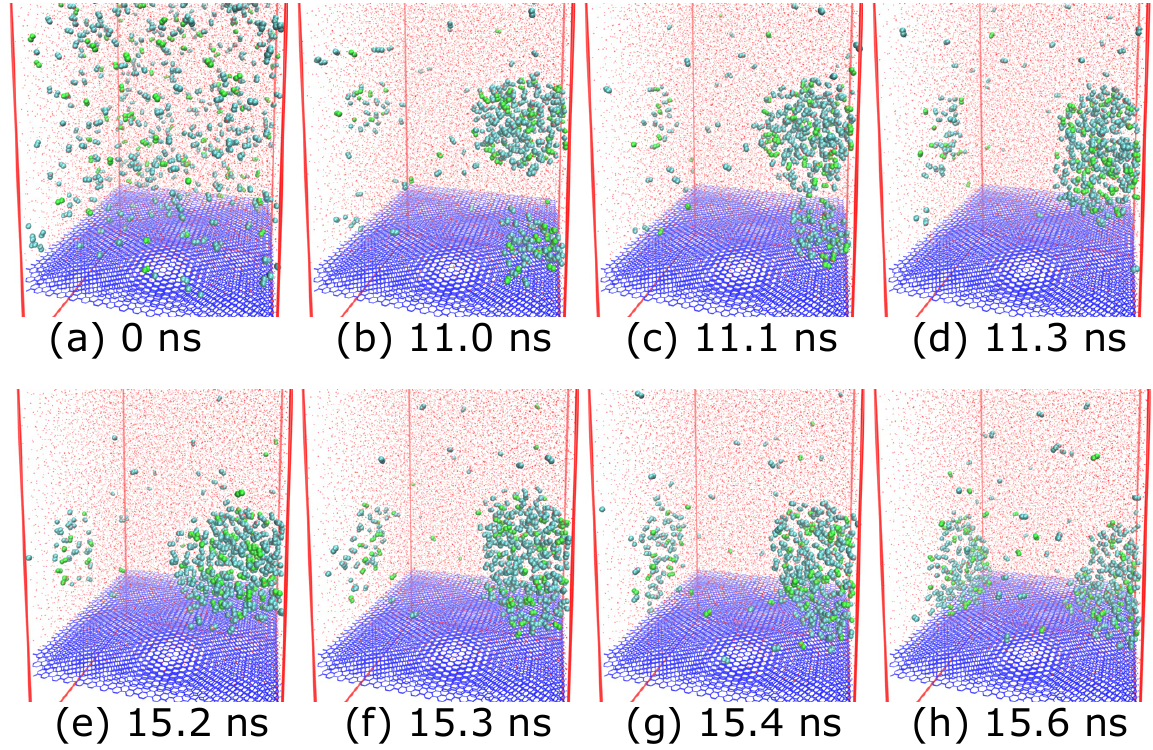}
    \end{center}
    \caption{\label{fig:4coalescence} Coalescence of two nanobubbles to form a large one and its subsequent spreading on the HOPG substrate illustrated with snapshots at different times as indicated. Two bubbles form initially in this system, a small one on the substrate and a large one in the bulk. The large nanobubble eventually merges with the small one on the HOPG substrate, in this way underlining the crucial energetic contribution of the substrate in the formation process. The domain is periodic in the $x$ and $y$ directions. A movie for the whole process is available as Supplementary Material Video 2.}
\end{figure}

\begin{figure}
    \begin{center}
    \includegraphics[width=\columnwidth]{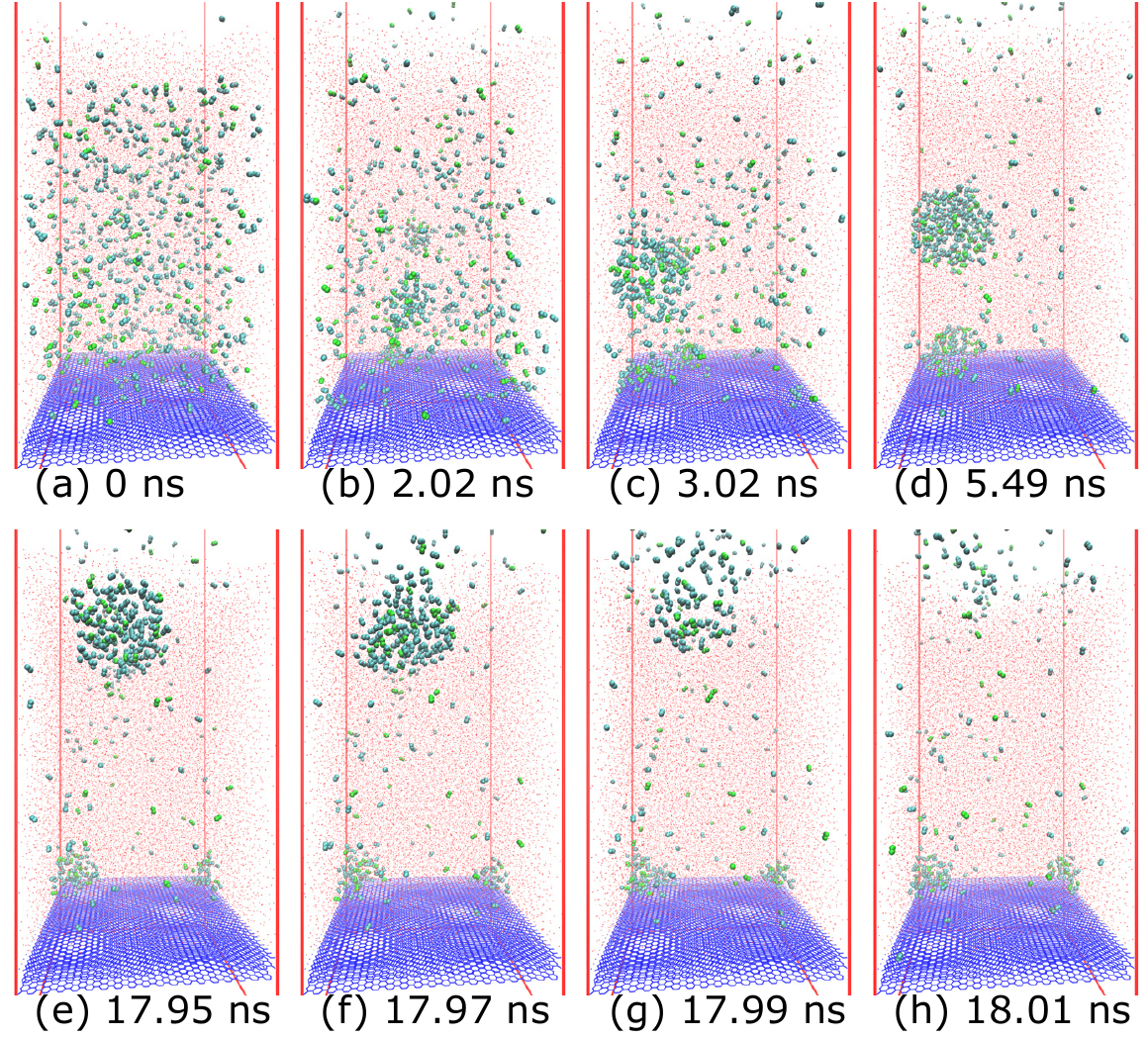}
    \end{center}
    \caption{\label{fig:5pop} Pop of a nanobubble towards the top water surface illustrated with snapshots at different times as indicated. A nanobubble produced in the bulk water phase approaches the top water surface, and pops immediately. Subsequently, all the gas molecules escape from the bubble to the open space. The domain is periodic in the $x$ and $y$ directions. A movie for the whole process is available as Supplementary Material Video 3.}
\end{figure}

During the formation of nanobubbles, several clusters of air molecules may collide and coalesce, resulting in a larger nanobubble (See Figure \ref{fig:4coalescence}). Some clusters of air molecules could also form in the liquid phase far from the substrate. These clusters exhibit a random (Brownian-like) motion, and pop immediately once they reach the air--water interface (see Figure \ref{fig:5pop}). In contrast, nanobubbles in contact with the substrate remain stable, in this way underlining the key role of the substrate in providing further stability to the nanobubbles.

\begin{figure}[tb]
    \begin{center}
    \includegraphics[width=\columnwidth]{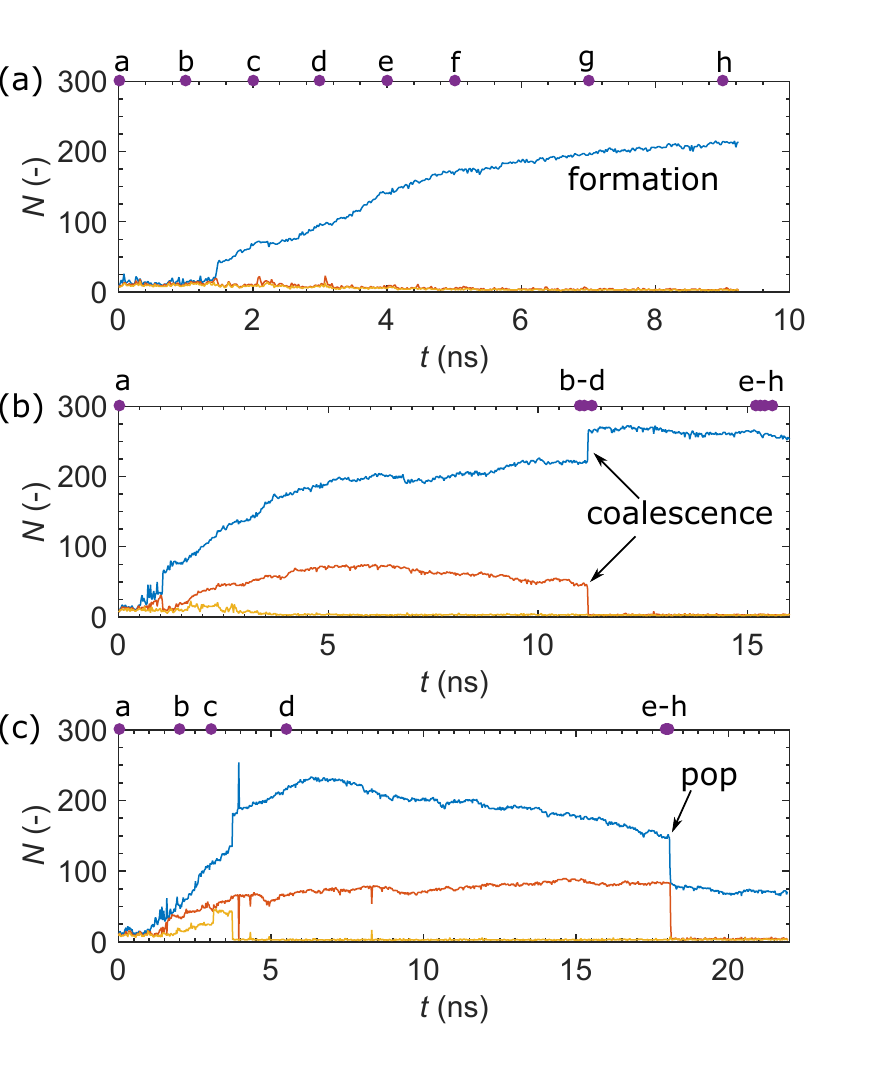}
    \end{center}
    \caption{\label{fig:growthCurve} Time evolution of the three largest clusters of air molecules (in terms of number of molecules) in the domain for three typical systems. (a) Formation of a nanobubble shown in Figure \ref{fig:6formation}, (b) coalescence of two large clusters shown in Figure \ref{fig:4coalescence}, (c) pop of the largest cluster shown in Figure \ref{fig:5pop}. Labels a--h on the axis of the plots correspond to the snapshots a--h in Figures \ref{fig:6formation}, \ref{fig:4coalescence}, and \ref{fig:5pop}, respectively.}
\end{figure}

The formation process realized by the aggregation of air molecules depends on the degree of saturation of the system. For the simulated systems in this study, the size of a typical nanobubble, during the formation process, increases rapidly within 5 ns until it reaches a plateau (see Figure \ref{fig:growthCurve}a) and forms nanobubbles of about 200 air molecules. In experiments, this process would require more time due to the lower concentration of air molecules and the larger size of the nanobubbles than that in the simulations.
The coalescence and the pop of clusters of air molecules happen in similar time scales (see Figures \ref{fig:growthCurve}b and c for coalescence and pop respectively). The coalescence time scale depends on the average distance between clusters, while the pop time scale depends on the average distance between the clusters and the water surface. Since these distances are functions of the initial air concentration and the air cluster size, they are much larger in experiments and consequently the time scales of the coalescence and the pop processes are also much larger in experiments than those in the simulations.

\begin{figure}
    \begin{center}
    \includegraphics[scale=0.6]{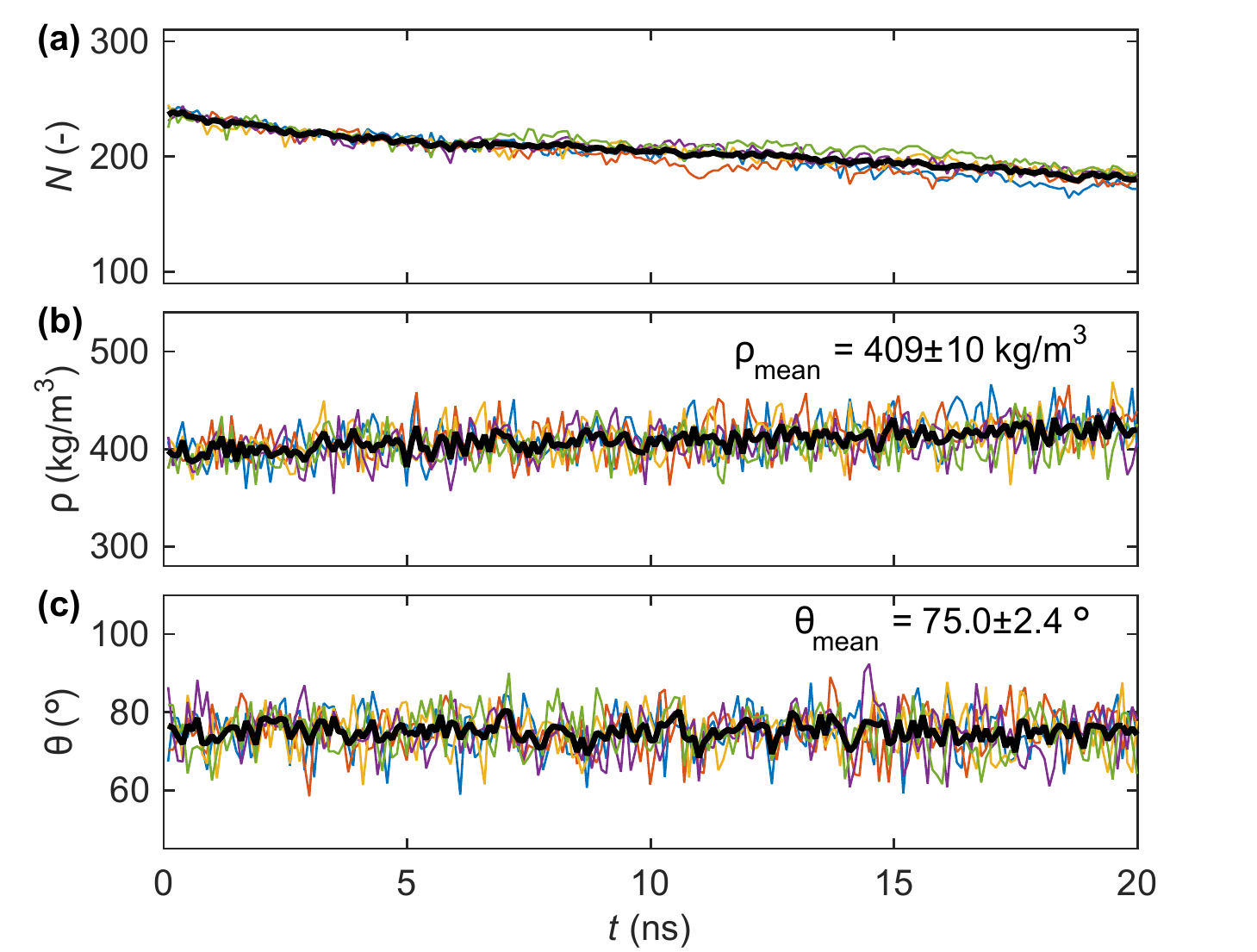}
    \end{center}
    \caption{\label{fig:bubbleProperty} Properties of the surface nanobubbles. (a) Size of nanobubbles in term of the number of molecules; the bubbles became smaller as air molecules is dissolved in water; (b) density of nanobubbles; (c) contact angle of nanobubbles. Curves with different colors represent simulations with different initial configurations (i.e.\ different random positions of air molecules), and the black curves show the mean values of five simulations.}
\end{figure}

During the formation of a nanobubble, the fast aggregation of air molecules indicates a collective process, where air molecules also reach diffusively the HOPG substrate. Then the nanobubble gradually dissolves at a pace that depends on its maximum size in terms of the number of contained air molecules (see Figure \ref{fig:bubbleProperty}a). In contrast to the formation process, the gradual dissolution of nanobubbles indicates a non-collective process, where the average number of air molecules in nanobubbles smoothly decays.
Interestingly, different simulations of nanobubbles with the same size exhibit the same behavior as they dissolve. The dissolution time of a nanobubble progresses linearly. For example, for a nanobubble of 245 air molecules shown in Figure \ref{fig:bubbleProperty}a, it is about 96 ns. Large nanobubbles will have longer life time, but could not be simulated using the MD method because of the high demand in CPU time. Moreover, longer lifetime of nanobubbles in experiments has been mainly attributed to the pinning of contact lines on defects of substrates \cite{Liu2014,Liu2014b,Guo2015,Guo2016}, which is not considered in the current study on smooth substrates. Large nanobubbles are dominated by the bulk interactions among air molecules, instead of the unfavorable interactions between water and air molecules at their interface, which tends to dissociate a small nanobubble. Hence, for a small nanobubble, these interfacial interactions lead to their gradual dissolution and the diffusion of air molecules to the top gas phase, where the system eventually reaches its global thermodynamic equilibrium state.

\subsection{Properties of nanobubbles}
To understand the behavior of nanobubbles on substrates, we sought for additional information by measuring different properties of the nanobubbles during our simulations. This would also assist in explaining the intriguing stability of nanobubbles. From our analysis, we find that the properties, such as density, contact angle, and surface tension, are significantly altered by the presence of the substrate and the size of the nanobubbles.

\textbf{Density of nanobubbles}
We find that nanobubbles are much denser than a typical gas phase, and there densities are rather close to typical densities of liquids.
In particular, the density of air molecules in nanobubbles is $409\pm 10$ kg/m$^3$ (see Figure \ref{fig:bubbleProperty}), of the order of the density of liquids (e.g., the densities of liquid nitrogen and liquid oxygen are 808 and 1141 kg/m$^3$ respectively, while the densities of nitrogen and oxygen at atmospheric state are 1.25 and 1.43 kg/m$^3$ respectively).
A more detailed analysis of the simulation data reveals that the average distance between air molecules is in the range of the van der Waals interactions indicating that air molecules are condensed in the nanobubble phase.
Hence, instead of the standard collisions in gases, van der Waals interactions dictate the behaviors of the air molecules in the condensed phase of a nanobubble. Therefore, air molecules in nanobubbles are dominated by the potential interactions instead of the kinetic energy, providing in this way further stability for nanobubbles.
Many experimental measurements could previously only confirm that nanobubbles entirely consist of air molecules \cite{Lohse2015}. In this study, we are able, for the first time, to provide an estimate for the density of nanobubbles, which was previously unknown.

\textbf{Contact angle of nanobubbles}
From the contact angle analysis for nanobubbles, we find that the contact angle measured in the case of nanobubbles is different from that measured in the respective case of nanodroplets, which we simulated for comparison (see Supplementary Material). In the case of an aqueous nanodroplet on the same HOPG substrate and in the air environment as in the case of nanobubbles, the contact angle is about 65$^\circ$, which is smaller than the contact angle in the nanobubble scenario, i.e., about 75$^\circ$ as shown in Figure \ref{fig:bubbleProperty}c (both contact angles are measured from the water side). This shows that the contact angle is not only a property of the substrate, but also depends on the configuration of the system (e.g., the ratio of water to air molecules). Moreover, in small systems the substrate interacts with the whole nanobubble/nanodroplet to affect the interface shape, unlike in large scale systems where the contact angle is a local behavior near the contact line. Nanobubbles are dominated by the interfacial interactions between air--water interface and the substrate, rather than the interaction of the air molecules in the bulk with the substrate. In agreement with previous findings \cite{Theodorakis2015}, the contact angle of nanodroplets depends on their sizes up to a system-dependent threshold. Our conclusions are consistent with experimental findings on the contact angles of nanobubbles, which argue that contact angles of nanobubbles are much larger than that of bubbles at large scales \cite{Lohse2015}.

\textbf{Surface tension}
Our analysis shows that the surface tension of nanobubbles is smaller than that in the corresponding bulk systems. We simulated the interfacial tension of the water--air interface of the nanobubbles at the atmospheric density and at the nanobubble density, which represent the surface tensions for bulk systems and for nanobubbles, respectively. The results reveal that the interfacial tension for the nanobubble case ($50.7 \pm 1.6$ mN/m) is smaller than that for water--air interface at atmospheric conditions ($57.6 \pm 3.0$ mN/m), where the latter value is in agreement with the model prediction from MD simulations \cite{Vega2007,Zhang2015SurfTen}. The relatively smaller surface tension of nanobubbles than that in the equivalent system under atmospheric conditions indicates the strong effect of the gas phase. This could be attributed to the higher density of air molecules in nanobubbles. The surface tension results from the net force produced by the greater attraction between water molecules than that between water and air molecules. A higher density of air can reduce the energetic penalty for the formation of the liquid--vapor interface.

Overall, our investigation of the density, contact angle, and surface tension of nanobubbles has underlined their dependence on the size of nanobubbles and the presence of the substrate. The air molecules in nanobubbles are significantly condensed and constraint into a small space, which affects the shape and contact angle of the nanobubbles. The surface tension is also suppressed due to the compression of the nanobubble. These are factors that contribute to the unusual stability of nanobubbles, which adopt an ``optimum'' metastable state with an increased density and contact angle, and a reduced surface tension.

\section{Conclusions}

By using all-atom simulations, the present work has clarified the role of the substrate in the formation/dissolution and properties of surface nanobubbles on HOPG substrates, which are challenging to investigate experimentally, but are important in understanding the intriguing stability of nanobubbles. The formation process is driven by an initial nucleation of air molecules and the subsequent coalescence of air clusters. The clusters favorably form on the substrate and are relatively more stable on the substrate than those in the bulk, which either move randomly to the substrate and spread or move to the liquid surface and pop immediately. Hence, the presence of the substrate is a crucial element for the enhanced stability of nanobubbles. The nanobubbles are found to consist of a high-density gaseous phase, an increased contact angle and a reduced surface tension with respect to equivalent bulk systems. These properties are also found to increase the stability of nanobubbles by providing an ``optimum'' metastable state.

Our results also demonstrate the possibility of addressing some fundamental questions regarding the field of nanobubbles by molecular dynamics simulations based on atomistic force fields. There are many open questions worthy of numerical and experimental investigation. The current approach could be extended to the study of substrate wettability, contamination \cite{Ducker2009}, and pinning \cite{Liu2014b,Guo2015,Guo2016}, which deserve careful consideration in order to understand the behaviors of nanobubbles. Direct measurement of the formation and the properties of nanobubbles, such as density, surface tension, and contact angle, would be helpful. We anticipate that our study will advance this field by providing insights into the physics of nanobubbles which are currently experimentally inaccessible. The findings of this work will be valuable for the understanding of many relevant phenomena in nature, such as nucleation in boiling \cite{Kim2009Boiling} and cavitation \cite{blake1987cavitation}, because nanobubbles could act as nucleation sites in these processes. The findings of this study can also be used in a wide range of potential applications of nanobubbles, such as material synthesis, microfluidics and nanofluidics, advanced diagnostics, and drug delivery.

\section*{Acknowledgements}
This work is supported by National Natural Science Foundation of China (Grant No.\ 51676137) and the National Science Centre, Poland (Grant No.\ 2015/19/P/ST3/03541).
This project has received funding from the European Union?¡¥s Horizon 2020 research and innovation programme under the Marie Sk{\l}lodowska-Curie grant agreement No.\ 665778.
{\footnotesize
\bibliography{nanobubble}\balance}

\end{document}